\documentclass{elsart}

\newcommand{\R}{{\mathbf R}} \newcommand{\N}{{\mathbf N}}
 
  \def\C{{\mathbf C}}

\newcommand{\wt}{\widetilde }

\renewcommand{\epsilon}{\varepsilon } 
 
\renewcommand{\rho}{\varrho } 
 
\renewcommand{\phi}{\varphi }
\renewcommand{\a}{\alpha }

\def\rs{\right>}
\def\lg{\left|}

\newtheorem{theorem}{Theorem}

\begin {document}
\begin{frontmatter}
 \title{From Monte Carlo to Quantum Computation}

\author {Stefan Heinrich}
\address{Fachbereich Informatik\\
Universit\"at Kaiserslautern\\
D-67653 Kaiserslautern, Germany\\
e-mail: heinrich@informatik.uni-kl.de\\
homepage:\\http://www.uni-kl.de/AG-Heinrich}

\date{}
\maketitle 
\begin{abstract}
Quantum computing was so far mainly concerned with discrete problems. 
 Recently, E. Novak and the author studied
quantum algorithms for high dimensional integration and 
 dealt with the question, 
which  advantages quantum computing can bring over classical deterministic
or randomized methods for this type of problem. 

In this paper we give a short introduction to the basic ideas of quantum computing
and survey recent results on high dimensional 
integration. We discuss connections to the Monte Carlo methology
and compare the optimal error rates of quantum algorithms
to those of classical deterministic and 
randomized algorithms.
\end{abstract}
\end{frontmatter}
{\large\bf Introduction}

One of the most challenging questions of today, in the overlap of computer science, 
mathematics, and physics, is the exploration of potential 
capabilities of quantum computers. Milestones were the algorithm of 
Shor \cite{Sho94}, who showed that quantum computers could 
factor large integers efficiently (which is widely believed to be infeasible on 
classical computers) 
and the quantum search algorithm of Grover \cite{Gro96}, which  
provides a quadratic speedup over deterministic and randomized classical algorithms of 
searching a database.

So far research was mainly concentrated on discrete problems like the above and 
many others one encounters in computer science.
Much less
is known about computational problems of analysis, including such 
typical field of application of Monte Carlo methods as high dimensional integration.
We seek to understand how well these problems can be solved
in the quantum model of computation (that is, on a -- hypothetical -- quantum computer)
and how the outcome compares to the efficiency of deterministic or Monte Carlo algorithms
on a classical (i.\ e.\ non-quantum) computer.

Abrams and Williams \cite{AW99} suggested first ideas about quantum integration
algorithms. 
A systematic study was begun by Novak \cite{Nov01}, who considered integration of  
functions from 
H\"older spaces.  This line of research was continued by the author \cite{Hei01a}, 
where quantum algorithms  for
the integration of $L_p$-functions and, as a key prerequisite,
for the
computation of the mean of $p$-summable sequences were constructed. In \cite{Hei01a}
a rigorous model of quantum computation for numerical problems was developed, as well.  
The case of integration of functions from Sobolev spaces is considered in \cite{Hei01b}, 
and more on the computation of the mean is presented in \cite{HN01b}.
These papers also established  matching lower bounds. 
A short survey of first results can be found in \cite{HN01a}. Path integration is
studied by Traub and Wo\'zniakowski in \cite{TW01}.

Combining these results with previous ones of information-based complexity theory
about the best possible ways of solving the respective problems deterministically
or by Monte Carlo on classical computers, we are now in a position to fairly well 
answer the question where quantum computation can provide a speedup in high 
dimensional integration and where not.
There are cases where quantum algorithms yield an exponential
speedup over deterministic algorithms and a quadratic speedup over randomized ones
(on classical computers).

Moreover, there is a close connection of quantum algorithms with Monte Carlo: 
While computations are carried out on superpositions of classical states of qubit systems,
and thus in high parallelism, 
the result can only be accessed through a measurement, which destroys the superposition
and outputs any one of the superposed states -- with a certain probability.
Thus, these algorithms are probabilistic, Monte Carlo, while, on the other hand, 
completely different laws govern the computation.
Nevertheless various Monte Carlo techniques can be put into use to
construct quantum algorithms (which then, combined with special quantum techniques,
outperform their classical counterparts).

We start with a brief  
 introduction to the ideas of quantum computing.
Then we consider the question, what quantum computers could do in numerical 
(Monte Carlo related) problems and
survey recent results on summation and high dimensional integration.
We discuss how to
use Monte Carlo methodology for the development of quantum algorithms.
Complexity issues and comparisons of the potential of deterministic, 
randomized and quantum algorithms are considered, as well.

For further reading 
on quantum computation we recommend the surveys by Aharonov \cite{Aha98},
Ekert, Hayden, and Inamori \cite{EHI00}, Shor \cite{Sho00}, and the monographs by 
Pittenger \cite{Pit99}, 
Gruska \cite{Gru99}, and Nielsen and Chuang \cite{NC00}. 
For notions and results in in\-for\-ma\-tion-based
complexity theory see the monographs by Traub, Wasilkowski, and 
Wo\'z\-nia\-kowski \cite{TWW88} and Novak \cite{Nov88}, and the survey 
Heinrich \cite{Hei93} of the randomized setting. 
\section{A Short Introduction to Quantum Computing}
\subsection{History}
The first ideas of using quantum devices for computation were expressed at the beginning
of the eighties by Manin \cite{Man80}, see also \cite{Man99}, and Feynman \cite{Fey82}.
They observed that simulating quantum mechanics on a classical computer is
extremly hard, probably infeasible, since it leads to differential equations
 whose dimensions are exponential in the number of system components.
To overcome this, they suggested the idea to simulate  quantum mechanics using 
quantum devices itself.

In 1985,  Deutsch \cite{Deu85} developed the rigorous theoretical basis of quantum computation
-- the model of a quantum Turing machine, which 
became the so far most serious and still standing
 challenge to the Turing-Church Thesis (the latter stating that, 
roughly, every reasonable physical computing device can 
be simulated with only polynomial
increase of resources on a classical Turing machine).

A breakthrough for quantum computing happened in 1994,  when Shor \cite{Sho94}
showed that efficient factorization of integers would be possible on a quantum computer, 
which in turn, would mean the possibility of breaking the foremost public key codes like the 
RSA cryptosystem. 
Another fundamental contribution was Grover's \cite{Gro96} discovery 
of an efficient
quantum search algorithm in 1996 (we give some further comments on both Shor's and Grover's 
algorithms later on).

 Since then we witness an explosion of efforts, broad research  
on quantum algorithms for all kinds of (mostly discrete) problems, 
on quantum cryptography, and quantum information theory. 
Physicists are intensively working on how to construct quantum computers,
that means, finding quantum mechanical systems that can be manipulated to fulfill the
abstractly proposed requirements. Systems with a few qubits are already
successfully realized in laboratories.

Important forerunners for the development of quantum algorithms
for Monte Carlo related, numerical problems were the work of
Boyer, Brassard, H{\o}yer, Mosca, and Tapp \cite{BBHT98}, \cite{BHMT00}
on quantum counting and the 
results of
Beals,  Buhrman, Cleve, Mosca and de Wolf \cite{BBCMW98} and 
Nayak and Wu \cite{NW99} on lower bounds. 
\subsection{Quantum Bit Systems}
In the classical physical world, in classical computation, a
bit is represented by two states of a physical system 
(e.g. charge or no charge) $\lg 0\rs$, $\lg 1\rs$. In the
(sub)atomic world, which is governed by the laws of quantum mechanics,
we have, along with the classical states $\lg 0\rs$, $\lg 1\rs$  
(such states could be, e.g., spin up or spin down of an electron),
also their superpositions:
$$\a_0\lg 0\rs+\a_1 \lg 1\rs\quad
(\a_0,\a_1\in\C, \,|\a_0|^2+|a_1|^2=1),$$
that is, the linear combination of the classical states
$\lg 0\rs$ and $\lg 1\rs$. The crucial problem however is that
if the system is in state
$\a_0\lg 0\rs+\a_1 \lg 1\rs$, we cannot access, 
measure $\a_0$ and $\a_1$ directly. Instead, a measurement destroys the 
superposition, causing the system to return to
state $\lg 0\rs$ with probability $|\a_0|^2$ and to 
state $\lg 1\rs$ with probability $|\a_1|^2$. 

The mathematical framework for a quantum bit, that is, 
a quantum system with two classical
basis states, is the 
two dimensional complex Hilbert space
$H_1:=\C^2$. Let 
$e_0,e_1$ be its unit vector basis.
Following  quantum mechanical notation, we write
$\lg 0\rs$ instead of $e_0$
and $\lg 1\rs$ instead of $e_1$. The unit sphere of $H_1$, i.e., the set of all
elements of norm 1, is the set of states of the qubit.

A system of $m$ interacting qubits ($m$-qubit system) is represented by the 
tensor product
$$
H_m:=\underbrace{H_1\otimes H_1\otimes\dots\otimes H_1}_{m},
$$
which is  the $2^m$-dimensional complex Hilbert space. It has the canonical basis
$$
e_{i_0}\otimes e_{i_1}\otimes\dots\otimes e_{i_{m-1}}\quad
(i_0,i_1,\dots,i_{m-1})\in \{0,1\}^m.
$$
We make further notational conventions:
$$
e_{i_0}\otimes e_{i_1}\otimes\dots\otimes e_{i_{m-1}}
=:\lg i_0\rs\lg i_1\rs\dots\lg i_{m-1}\rs
=:\lg i\rs,
$$
where $i:=(i_0i_1\dots i_{m-1})_2:=\sum_{k=0}^{m-1}i_k 2^{m-1-k}$.
The vectors 
$\lg i\rs\; (i=0,\dots,2^m-1)$ are called classical states,
while a general state has the form
$$
\lg \psi\rs=\sum_{i=0}^{2^m-1}\a_i\lg i\rs\qquad 
\left(\sum_{i=0}^{2^m-1}|\a_i|^2=1\right).
$$ 
As in the case of a single qubit, measuring an $m$-qubit system in a superposition
state $\lg\psi\rs$ results in
one of the classical states $\lg i\rs$ with probability 
$|\a_i|^2\quad (i=0,\dots,2^m-1)$. 
So a state is a linear combination of all possible classical states,
the coefficients giving the probability that after measurement the system
moves to this state.\\\\
One more notational convention: if $\lg \cdot\rs $ contains a number inside,
or a symbol which is used to denote such a number, 
we mean the canonical basis vector corresponding to this number,
if $\lg \cdot\rs $ contains just a general symbol, like $\lg \psi\rs $,
we mean any vector of $H_m$ (this should be clear from the context).

\subsection{Quantum Computation}
How to use such systems for computing?
To make this clear, let us first consider an example of a classical computation
-- the addition of two $m$-bit numbers, which we write as follows:
\begin{eqnarray*}
&&\lg i_0\rs\dots\lg i_{m-1}\rs\lg j_0\rs\dots\lg j_{m-1}\rs
\lg 0\rs\,\,\dots\,\lg 0\rs\\
&&\qquad\qquad\qquad\quad\downarrow\\
&&\lg i_0\rs\dots\lg i_{m-1}\rs\lg j_0\rs\dots\lg j_{m-1}\rs
\lg k_0\rs\dots\lg k_{m}\rs
\end{eqnarray*}
This computation is realized using circuits of classical gates
({\tt and, or, not, xor}) in the usual way: add the last bits, 
 then the second last plus the carry bit 
etc. Let us emphasize here:
{\it Classically, we add two numbers at a time.}\\
\\
How to operate $m$-qubit quantum systems?
Which operations are allowed?
Schr\"odinger's equation implies: all evolutions of a quantum system 
must be represented by  unitary transforms of $H_m$. Here is the starting point: 
\\\\
{\it Quantum computing assumes that we are able to perform a number of 
elementary unitary transforms
(quantum gates) on the system.}
\\\\
What are these operations? Let us consider one standard set $\mathcal{G}_m$ of them.
First we  describe the one-qubit gates 
-- these are gates that manipulate only one component of 
the tensor product
$H_m=H_1\otimes H_1\otimes\dots\otimes H_1$.
The Walsh-Hadamard gate $W: H_1\to H_1$ is defined by
$$
W\lg 0\rs = \frac{\lg 0\rs+\lg 1\rs}{\sqrt{2}}\qquad
W\lg 1\rs = \frac{\lg 0\rs-\lg 1\rs}{\sqrt{2}}
$$
(the values on the basis vectors define the unitary transform uniquely). Its action on
the $j$-th component of $H_m$ is then given by the unitary operator
$$
W_m^{(j)}=Id\otimes\dots\otimes Id\otimes \underbrace{W}_{j}
\otimes Id\otimes\dots\otimes Id,
$$
where $Id$ stands for the identity operator on $H_1$.
For a real parameter 
$0\le \theta<2\pi$ the phase shift $P_\theta: H_1\to H_1$ is  defined as 
$$
P_\theta\lg 0\rs = \lg 0\rs\qquad
P_\theta\lg 1\rs = e^{\imath \theta}\lg 1\rs.
$$
We define $P_{\theta,m}^{(j)}$ in the respective way.
Next we consider two-qubit gates -- they manipulate any chosen two components of 
$H_1\otimes H_1\otimes\dots\otimes H_1$. The
 quantum xor gate (also called  controlled-not gate) 
$X:H_1\otimes H_1\to H_1\otimes H_1$ is given by
\begin{eqnarray*} 
X\lg 0\rs\lg 0\rs= \lg 0\rs\lg 0\rs&&\\
X\lg 0\rs\lg 1\rs= \lg 0\rs\lg 1\rs&&\\
X\lg 1\rs\lg 0\rs= \lg 1\rs\lg 1\rs&&\\
X\lg 1\rs\lg 1\rs= \lg 1\rs\lg 0\rs&&
\end{eqnarray*}
That is, if the first bit is zero, nothing happens to the second, 
and if the first is one, the second is negated (controlled not).
We can look at this gate also as follows: the xor of the two bits replaces
the second bit. Denote by $X_m^{(k,\ell)}:H_m\to H_m$ the unitary operator
given by applying $X$ to the $k$-th and $\ell$-th component, that is,  
$$
X_m^{(k,\ell)}\lg i_0\rs\dots\lg i_{k}\rs
\dots\lg i_{\ell}\rs\dots\lg i_{m-1}\rs\\
=\lg i_0\rs\dots\lg y\rs
\dots\lg z\rs\dots\lg i_{m-1}\rs,
$$
where
$$ 
\lg y\rs\lg z\rs:=X\lg i_{k}\rs\lg i_{\ell}\rs.
$$
Now we define 
$$
\mathcal{G}_m=\left\{W_m^{(j)},\, P_{\theta,m}^{(j)},\, X_m^{(k,\ell)}\: :
\: 0\le j,k\ne \ell\le m-1,\,0\le \theta<2\pi\right\}.
$$ 
 The following two results can be found, e.g., in \cite{NC00}.
\begin{theorem}
 The set $\mathcal{G}_m$ is a
universal system of gates -- each unitary transform of $H_m$ can be
represented as a finite composition of elements of $\mathcal{G}_m$
 (up to a complex scalar factor).
 \end{theorem}

The set  $\mathcal{G}_m$ is still an infinite set. Now consider the following finite
subset:
$$
\mathcal{G}_m^0=\left\{W_m^{(j)},\, P_{\pi/4,m}^{(j)},\, X_m^{(k,\ell)}\: :
\: 0\le j,k\ne \ell\le m-1\right\}.
$$
\begin{theorem} 
The set  $\mathcal{G}_m^0$ forms an approximately 
universal system of gates -- each unitary transform of $H_m$ can be
approximated in the operator norm to each precision by a finite 
composition of  elements of $\mathcal{G}_m^0$
(again up to a complex factor).
\end{theorem}
So once we can implement these gates we can do all unitary transforms
(of course, the efficiency of such an approximation is still an issue, 
we will come back to that later on).
Physicists are working on implementations of these gates in various quantum systems 
such as photons, trapped ions, magnetic resonance systems etc. Let me emphasize some 
crucial points: \\
1. These gates can transform classical states into superpositions. For
example, the Hadamard gate applied to the first and then to the second qubit:
$$
\lg 0\rs\lg 0\rs\longrightarrow
\frac{1}{2}\left(\lg 0\rs\lg 0\rs+\lg 0\rs\lg 1\rs+\lg 1\rs\lg 0\rs
+\lg 1\rs\lg 1\rs\right).
$$
2. They also act on superpositions. For instance:\\  
2.1 The quantum xor:
\begin{eqnarray*}
&\a_0\lg 0\rs\lg 0\rs+\a_1\lg 0\rs\lg 1\rs+\a_2\lg 1\rs\lg 0\rs+\a_3\lg 1\rs\lg 1\rs&\\
&\downarrow&\\
&\a_0\lg 0\rs\lg 0\rs+\a_1\lg 0\rs\lg 1\rs+\a_2\lg 1\rs\lg 1\rs+\a_3\lg 1\rs\lg 0\rs&
\end{eqnarray*}
2.2 Quantum addition of binary numbers:
\begin{eqnarray*}
&\sum\a_{ij}\lg i_0\rs\dots\lg i_{m-1}\rs\lg j_0\rs\dots\lg j_{m-1}\rs
\lg 0\rs\,\,\dots\,\lg 0\rs\,\,\,\,&\\
&\quad\downarrow&\\
&\sum\a_{ij}\lg i_0\rs\dots\lg i_{m-1}\rs\lg j_0\rs\dots\lg j_{m-1}\rs
\lg k_0\rs\dots\lg k_{m}\rs&
\end{eqnarray*}
{\it That is, in the quantum world, we add all possible binary $m$-digit numbers
in parallel}
(assuming that all $\a_{ij}\ne 0$). So, is a quantum computer an ideal 
parallel computer, with exponentially 
many processors? Not exactly, it is not that easy!
We cannot access all components of the superposition.
We have to measure, thus destroying all results but one!
We would get
$$\lg i_0\rs\dots\lg i_{m-1}\rs\lg j_0\rs\dots\lg j_{m-1}\rs
\lg k_0\rs\dots\lg k_{m}\rs$$
with probability $|\a_{ij}|^2$. In this sense a quantum computer is
like a black box -- we can manipulate it, but we cannot ''look into it''.
Anyway, we have reached the quantum model of computation
(the general way a quantum algorithm should look like):\\
\\
{\bf Quantum model of computation:}
\[
\begin{array}{ll}
\mbox{starting state:}& \lg i_0\rs \in H_m,\:\mbox{ a classical state}\\
\mbox{computation:}& U_1,\dots,U_n \in \mathcal{G}_m\\
&\lg i_0\rs\to U_1\lg i_0\rs\to U_2 U_1\lg i_0\rs \dots\\
&\to\lg \psi\rs:=U_n U_{n-1}\dots U_2 U_1\lg i_0\rs\\
\mbox{measurement:} & \lg \psi\rs=\sum_{i=0}^{2^m-1}\a_i\lg i \rs
\\
&\to
 \lg i \rs\,\mbox{with probability}\ |\a_i|^2\\
\mbox{repeat}\end{array}
\]
The cost of the algorithm is defined to be the number of gates $n$.
We may allow for a classical computation after each measurement to determine
the next starting state, and, after the last measurement, the final
result (see \cite{Hei01a} for a formal approach).
Then, speaking about the cost of the algorithm, we also have to take 
into account the needed classical gates. A quantum algorithm is said to solve
a given problem if it produces a solution with probability $\ge 3/4$ (any
fixed constant $\gamma>1/2$ would do, since the success probability can always be
increased by suitable repetitions).
\subsection{Computational Power of Quantum Computers}
What can we do with a quantum computer that we cannot do on a classical computer?
A simple answer related to Monte Carlo algorithms can be given as follows:
On classical computers, pseudo-random numbers (that is, in fact, fully 
deterministic numbers) are used instead of random numbers. A quantum computer would 
provide us with true randomness. It is simple to let a quantum computer act as a 
random number generator:\\\\
\mbox{\bf  Quantum random number generator}
\begin{eqnarray*}
&\lg 0\rs\dots\lg 0\rs&\\[.2cm]
&\downarrow&\mbox{(Walsh-Hadamard transform)}\\[.2cm]
&\displaystyle 2^{-m/2}\sum_{i=0}^{2^{m}-1}\lg i\rs&\\[.2cm]
&\downarrow&\mbox{(measurement)}\\[.2cm]
&\lg i\rs\,\,\mbox{with probability}\,\,\, 2^{-m}&
\end{eqnarray*}
So, on a quantum computer, we can implement Monte Carlo with 
true (physically based) randomness. Let us go one step further:
What can be done on a quantum computer that a {\it randomized} classical
computer (that is, a classical computer with access to true
randomness) cannot do? We already mentioned  
Shor's result \cite{Sho94}: An algorithm which, for a composite integer $N$, 
finds a nontrivial factor with cost $\mathcal{O}((\log N)^3)$. 
No polynomial in $\log N$ classical (deterministic or randomized) algorithm is known.
Recall that there are polynomial randomized prime number tests 
(see Solovay and Strassen \cite{Sol77},  and Rabin \cite{Rab80}), 
but they don't give the factors
when they find compositeness.

Although not very likely, it is not excluded that there are polynomial 
classical factoring algorithms (maybe even with cost $\mathcal{O}((\log N)^3)$).  
So formally, the above result does not strictly prove  the superiority of
quantum computing. Matters are different with Grover's result \cite{Gro96}. The
problem is the following: 
Given $f:\{0,\dots,N-1\}\to\{0,1\}$ with the property 
that there is a unique $i_0$ with $f(i_0)=1$,  
find this $i_0$. It is not difficult to see that
classical (deterministic or randomized) algorithms need cost
 $\Omega(N)$. Grover \cite{Gro96} produced a quantum algorithm that solves the problem 
with cost $\mathcal{O}(\sqrt{N})$. Here we have to extend the above model of 
computation by what is called a quantum query (quantum subroutine, quantum oracle), 
which is given by the unitary map $Q_f:H_m\to H_m$ defined as
$$
Q_f:\lg i\rs\lg y\rs \to \lg i\rs\lg y\oplus f(i)\rs
\quad(0\le i<2^{m-1},\,y\in \{0,1\}),
$$
where $\oplus$ is addition modulo 2, and we assumed $N=2^{m-1}$. 
Note that if the last qubit 
is set to zero, $Q_f$ is like a 
subroutine which writes the function value to the last entry.
This is what we essentially assume in the design of most numerical and 
Monte Carlo algorithms: the function we want to handle, e.g., integrate,
is given as a subroutine.
(In the quantum setting, addition modulo 2 guarantees 
that $Q_f$ is a bijection
on the classical states and thus, extends to a unitary operator.)
It can be shown that such a quantum query can always be implemented efficiently
(with only a constant factor increase of resources),
once we have a classical implementation for $f$.
A quantum computation is then still given by the scheme in section 1.3, but
with the difference that $U_i\in \mathcal{S}_m \cup \{Q_f\}$.
It is important to mention that the rest of the $U_i$ should not depend on $f$,
that is, the algorithm gets its information  about $f$ only through query 
calls. The cost is defined to be the number of query calls, multiplied by the 
number of qubits, plus the number of gates .

\section{Summation and Integration}
Now we consider numerical problems. Let us formulate the
general integration problem: Given a set
$D$, a measure $\mu$ on it, and a $\mu$-integrable function $f:D\to \R$, 
compute (approximate)
$$
\int_{D}f(t)d\mu(t).
$$
As its simplest instant, this includes computing the mean (and hence the sum)
of a finite sequence:
$$
D=\{0,\dots,N-1\}, \quad f:D\to \R,
\quad \mu(\{i\})=\frac{1}{N}$$
$$
\int_{D}f(t)d\mu(t)=\frac{1}{N}\sum_{i=0}^{N-1}f(i). 
$$
\begin{theorem}\label{theo:1}{\rm (Brassard, H{\o}yer, Mosca, and Tapp 
\cite{BHMT00})}
For all $n,N$, $n<N$, there is a quantum algorithm which computes the mean 
$\frac{1}{N}\sum_{i=0}^{N-1}f(i)$ for all sequences $f(i)$ with $f(i)\in \{0,1\}$
with cost $\wt{\mathcal{O}}(n)$ and error $\mathcal{O}(1/n)$.
\end{theorem}
We use the 
$\wt{\mathcal{O}}$ notation to indicate that we suppress possible 
logarithmic factors.
The result is easily extended to real valued $f(i)\in [-1,1]$.
We have to specify how a quantum algorithm can access such values of
$f$. Without loss of generality we suppose that $N$ is of the form $N=2^{m_1}$.
We assume that we have a quantum query (subroutine) 
providing essentially the first $m_2$ digits of  $f(i)$ for some suitable $m_2$.
More precisely, we can use the unitary mapping $Q_f$ on 
$H_m=H_{m_1}\otimes H_{m_2}$ defined as 
$$
Q_f: \lg i\rs \lg y\rs\to \lg i\rs \lg y\oplus [f(i)]_{m_2}\rs
\quad(0\le i<2^{m_1},\,0\le y<2^{m_2}).
$$
Here $\oplus$ is addition modulo $2^{m_2}$ and $[f(i)]_{m_2}$ is the 
$m_2$-bit integer $\lfloor 2^{m_2-1}(f(i)+1)\rfloor$ if $f(i)<1$ and 
$2^{m_2}-1$ if $f(i)=1$. 
(If we have to deal with functions taking values in
other intervals, say $[a,b]$, we use a similar encoding as a binary integer,
$[f(i)]_{m_2}=\lfloor 2^{m_2}(f(i)-a)/(b-a)\rfloor$ if $f(i)<b$ and 
$[f(i)]_{m_2}=2^{m_2}-1$ if $f(i)=b$.)
We say that a quantum algorithm computes a number with error $\epsilon$, 
if with probability $\ge 3/4$ the result of the algorithm is within $\epsilon$ 
of that number. 
For more details on queries for numerical problems and the error definition 
we refer to 
\cite{Hei01a}. Theorem \ref{theo:1} should be viewed under the
aspect of huge $N$ and moderate $n$.
For comparison, let us mention that for classical deterministic
algorithms the error at cost $n$ for $n<N/2$ is  $\Omega(1)$ (even $n<cN$ for 
any fixed $0<c<1$ suffices). In the
classical randomized setting we obtain $\Theta(1/\sqrt{n})$ 
(see \cite{HN01a} for details on the latter two statements).

How to treat integration? Define the H\"older classes $\mathcal{F}_d^{r,\rho}$ for   
$r \in \N_0$, $d\in \N$, and $0<\rho \le 1$ as
\begin{eqnarray*}
\mathcal{F}_d^{r,\rho} &=&
 \{ f \in C^r([0,1]^d)\, :\:  \Vert f \Vert_\infty \le 1,\\ 
  &&|\partial^\a f (x) - \partial^\a f (y) | \le |x - y|^\rho, 
   |\a|=r  \}.
\end{eqnarray*}
Here
$C^r([0,1]^d)$ stands for the set of $r$ times continuously 
differentiable functions on $[0,1]^d$, $\|\;\|_\infty$ denotes the supremum norm, 
$\a$ represents a multiindex, and $\partial^\a$ is the respective partial
derivative.

\begin{theorem}\label{theo:2}{\rm (Novak \cite{Nov01})}
For all $n$, there is a quantum algorithm which computes the integral 
of functions $f\in \mathcal{F}_d^{r,\rho}$ 
with cost $\wt{\mathcal{O}}(n)$ and error \\ $\mathcal{O}(n^{-(r+\rho)/d-1})$.
\end{theorem}
Because of its relation to Monte Carlo methodology it is interesting to
describe the idea behind the algorithm: We use the technique of 
separation of the main part 
(in other words, of introducing a control variate) to reduce integration
to summation in an optimal way. Let
$P_n$ be some interpolation operator with $n$ nodes $(\tau_i)_{i=0}^{n-1}
\in D=[0,1]^d$ suitable for $\mathcal{F}_d^{r,\rho}$ in the sense that it 
gives the optimal approximation order
$$
\sup_{t\in D}|f(t)-P_nf(t)|=\mathcal{O}(n^{-(r+\rho)/d}).
$$
Represent   
\begin{equation}
\label{A1}
\int_D f(t)dt= \int_D P_nf(t)dt+\int_D (f(t)-P_n f(t))dt.
\end{equation}
Clearly, 
$$
\int_D P_n f(t)dt=\sum_{i=0}^{n-1} c_if(\tau_i)
$$
is a quadrature, which can be computed exactly, classically, with
 $\mathcal{O}(n)$ cost. 
The second part is approximated suitably (see \cite{Nov01} for details) in the form
$$
\int_D (f(t)-P_n f(t))dt
\approx\frac{1}{N}\sum_{i=0}^{N-1}(f(t_i)-P_n f(t_i)),
$$
where $N$ may be much larger than $n$.
Finally we use quantum computation to approximate the mean on the right hand side. 
It can be shown that the order of the error is the product of the following
contributions:
$$
\underbrace{n^{-(r+\rho)/d}}_{\mbox{separation of main part\,\,}}
*\underbrace{n^{-1}}_{\mbox{\,\,quantum computation}}
$$
Note that if the last term of (\ref{A1}) were approximated by standard Monte 
Carlo, we would get the exponent $-(r+\rho)/d-1/2$.

Compare this result with the classical deterministic setting, where
the best we can achieve is an error $\mathcal{O}(n^{-(r+\rho)/d})$
at cost $\mathcal{O}(n)$ and with the 
classical randomized setting with optimal error $\mathcal{O}(n^{-(r+\rho)/d-1/2})$.
Look at the exponents when $(r+\rho)/d$ is small! It follows that the speedup over 
the deterministic setting
can be polynomial with arbitrarily large power. Note that in \cite{Hei01a}, Section
6, a class of functions of low smoothness is considered and it is proved that
for this class the gain of quantum over classical deterministic computation is
even exponential. A similar situation is observed in \cite{TW01} for path
integration.
 
A crucial assumption for the summation algorithm in Theorem \ref{theo:1}, and
thus also for Theorem \ref{theo:2}, is the uniform boundedness of the 
set of sequences (with bounds independent of $n$ and $N$). 
This raised the problem whether the quantum gain was tied to
that assumption. In particular, what happens in the case of square summability,
which is the most important case for Monte Carlo, 
that is, $f\in \mathcal{B}_2^N$ or $f\in \mathcal{W}_{2,d}^r$ 
(see the definitions below)? 
Will quantum computation retain its superiority or will Monte Carlo catch up?
Even more generally, what about $f\in \mathcal{B}_p^N$ or  
$f\in \mathcal{W}_{p,d}^r$ for $1\le p<\infty$?
Let us first look at the discrete problem.
Denote 
$$
\mathcal{B}_p^N=\left\{f\: :\:\frac{1}{N}\sum_{i=0}^{N-1} |f(i)|^p\le 1\right\}.
$$
The summation problem for this class was settled in \cite{Hei01a} 
(case (i) and (ii)) and in \cite{HN01b} (case (iii)). The latter 
 answered a question posed in \cite{HN01a}.
\begin{theorem}\label{theo:3}
Let $1\le p <\infty$. For all $n,N$, $n<N$, there is a quantum algorithm which 
computes the mean 
$\frac{1}{N}\sum_{i=0}^{N-1}f(i)$ for all sequences $f\in\mathcal{B}_p^N$
with cost $\wt{\mathcal{O}}(n)$ and error 
\[
\begin{array}{lll}
(i)\qquad&\mathcal{O}(n^{-1})&\quad \mbox{if}\quad 2\le p<\infty,\\
(ii)\qquad&\mathcal{O}(n^{-2+2/p})&\quad \mbox{if}\quad 1\le p<2 
\,\,\mbox{and}\,\, n<\sqrt{N}, \\
(iii)\qquad&\mathcal{O}\left(n^{-2/p}N^{2/p-1}\right)
&\quad \mbox{if}\quad 1\le p<2 \,\,\mbox{and}\,\,\sqrt{N}\le n<N.
\end{array}
\]
\end{theorem}
Again, a comparison to the two classical settings might be illustrative (see
\cite{HN01a} for details): 
In the classical deterministic case, for $n<N/2$, nothing better than $\Omega(1)$
can be obtained, in the classical randomized case the optimal rates are  
$\Theta(n^{-1/2})$ if $2\le p<\infty$, and
$\Theta(n^{-1+1/p})$ if $1\le p<2$.
Now we consider integration in the
Sobolev classes $\mathcal{W}_{p,d}^r$, which are defined by 
$$
\mathcal{W}_{p,d}^r=
 \{ f\in L_p([0,1]^d) \,  : \,   \Vert \partial^\a  f \Vert_{L_p}  \le 1,
 \ |\a| \le r   \},  
$$
where $r \in \N$, $1 \le p \le \infty$, and $\partial^\a$ is the weak partial 
derivative. The following result is proved in \cite{Hei01b}. It
answers another question from \cite{HN01a}.
The approach consists of a
new discretization technique, by which one can
derive (optimal) integration algorithms for $\mathcal{W}_{p,d}^r$ from
(optimal) summation algorithms for  $\mathcal{B}_p^N$.

\begin{theorem}\label{theo:5} 
Let $1\le p <\infty$, $r,d\in \N$, $r/d>1/p$ (Sobolev embedding condition). 
For all $n$, there is a quantum algorithm which computes the integral 
of functions $f\in \mathcal{W}_{p,d}^r$  
with cost $\wt{\mathcal{O}}(n)$ and error 
$\mathcal{O}(n^{-r/d-1})$.
\end{theorem}
In the classical deterministic setting the optimal rate
is $\mathcal{O}(n^{-r/d})$, while in the 
classical randomized setting we have 
$\mathcal{O}(n^{-r/d-1/2})$ if $2\le p<\infty$, and 
$\mathcal{O}(n^{-r/d-1+1/p})$ if $1\le p<2$
(see \cite{HN01a}, \cite{Hei93}, and the references therein). 
The quantum rate for $1\le p<2$ comes as a surprise. After previous results one
was tempted to conjecture that the quantum setting could reduce the exponent
of the classical randomized setting by at most 1/2. 
Now we see ($p=1$) that there can even be a reduction by 1.
 
Are these results about quantum algorithms optimal? In other words,
is it possible to improve the rates by other, better quantum algorithms?  
To verify optimality, we have to establish lower bounds valid
for all possible quantum algorithms. It turns out that all 
the results about summation and integration presented here are optimal 
(up to logarithmic factors, at least).
The following was the first matching lower bound for summation.
It shows that the rate in Theorem \ref{theo:1} is optimal. 
\begin{theorem}\label{theo:6}{\rm (Nayak and Wu \cite{NW99})}
There are constants $c_1,c_2>0$ such that for all $n,N$ with 
$n<c_1 N$ the following holds:
Each quantum algorithm which computes the mean 
$\frac{1}{N}\sum_{i=0}^{N-1}f(i)$ for all sequences $f(i)$ with $|f(i)|\le 1$
using at most $n$ quantum queries has error not smaller than $c_2/n$.
\end{theorem}
Using their technique and methods of infor\-mation-based complexity theory, 
the following can be shown (see \cite{Nov01}, \cite{Hei01a}, \cite{HN01b}, 
\cite{Hei01b}).
\begin{theorem}\label{theo:8} The rates established in
Theorem \ref{theo:2} (integration in $\mathcal{F}_d^{r,\rho}$),
Theorem \ref{theo:3} (mean of sequences in $\mathcal{B}_p^N$), and
Theorem \ref{theo:5} (integration in $\mathcal{W}_{p,d}^r$)
are optimal (up to logarithmic factors).
\end{theorem}
In the following table we summarize the results. The respective entries
give the optimal rates at cost $n$, constants and logarithmic factors are suppressed.
The constant $0<c<1$ in the first column does not depend on $n$ and $N$.
\[\hspace*{-.5cm}
\begin{array}{l|l|l|l}
& \ \mbox{deterministic}\  & \, \mbox{random}\, & \, \mbox{quantum}\,\\ \hline 
\mathcal{B}_p^N,\,2\le p\le \infty,\,n<cN &\, 1& \, n^{-1/2}        & \, n^{-1}\\
\mathcal{B}_p^N,\,1< p<2,\,
n<\sqrt{N}& \, 1&\,  n^{-1+1/p} & \,  n^{-2+2/p}  \,\, 
\\
\mathcal{B}_p^N,\,1< p<2,\,
\sqrt{N}\le n<cN & \, 1&\,  n^{-1+1/p} & \,  n^{-2/p}N^{2/p-1}  \,\, 
\\
\mathcal{B}_1^N,\,n<\sqrt{N} & \, 1&\,  1 & \,  1  \,\, 
\\
\mathcal{B}_1^N,
\sqrt{N}\le n<cN & \, 1&\,  1 & \,  n^{-2}N \,\, 
\\
\mathcal{F}_d^{r,\rho} &\, n^{-(r+\rho)/d}  &\,  n^{-(r+\rho)/d-1/2 \  } 
& \,  n^{-(r+\rho)/d-1}\\
\mathcal{W}_{p,d}^r,\,2\le p\le \infty &\, n^{-r/d } & \, n^{-r/d -1/2 }  
& \, n^{-r/d -1}\\
\mathcal{W}_{p,d}^r,\,1< p<2 &\, n^{-r/d } & \, n^{-r/d -1 + 1/p }
&  n^{-r/d -1}\\ 
\mathcal{W}_{1,d}^r &\, n^{-r/d } & \, n^{-r/d }
&  n^{-r/d -1}\\ 
\end{array}
\]


\begin{thebibliography}{}
\bibitem{Aha98}
D. Aharonov (1998): 
Quantum computation -- a review. In: 
Annual Review of Computational Physics, 
World Scientific, volume VI, ed. Dietrich Stauffer, see also
http://arXiv.org/abs/quant-ph/9812037.

\bibitem{AW99}
D.~S. Abrams and C.~P. Williams (1999):
 Fast quantum algorithms for numerical integrals and stochastic
  processes.
 Technical report, http://arXiv.org/abs/quant-ph/9908083.

\bibitem{BBCMW98}
R.~Beals,  H.~Buhrman, R.~Cleve, M.~Mosca, and R.~de Wolf (1998):
 Quantum lower bounds by polynomials. 
 Proceedings of 39th IEEE FOCS, 352-361, see also
 http://arXiv.org/abs/quant-ph/9802049.

\bibitem{BBHT98}
M.~Boyer, P.~Brassard, P.~H{\o}yer, and A.~Tapp (1998):
\newblock Tight bounds on quantum searching. 
 Fortschritte der Physik {\bf 46}, 493 -- 505,
see also http://arXiv.org/abs/quant-ph/9605034.

\bibitem{BHMT00}
G.~Brassard, P.~H{\o}yer, M.~Mosca, and A.~Tapp (2000):
\newblock Quantum amplitude amplification and estimation.
\newblock Technical report, http://arXiv.org/abs/quant-ph/0005055.

\bibitem{BHT98}
G.~Brassard, P.~H{\o}yer, and A.~Tapp (1998):
\newblock Quantum counting.
\newblock  Lect. Notes in Comp. Science {\bf 1443}, 820 -- 831, see also 
  http://arXiv.org/abs/quant-ph/9805082.
  
  
\bibitem{Deu85}
 D.\ Deutsch (1985):
Quantum theory, the Church-Turing principle and the universal quantum computer.
Proc. R. Soc. Lond., Ser. A 400, 97-117.

\bibitem{EHI00}
A. Ekert, P. Hayden, and H. Inamori (2000): Basic concepts in quantum computation.
See http://arXiv.org/abs/quant-ph/0011013.

\bibitem{Fey82}
R.\ Feynman (1982): Simulating physics with computers. Int. J. Theor. Phys. 
{\bf 21}, 467--488.

\bibitem{Gro96} 
L. Grover (1996):
A fast quantum mechanical algorithm for database search. 
Proc. 28 Annual ACM Symp. on the Theory of Computing, 212--219, ACM Press New York.  
See also http://arXiv.org/abs/quant-ph/9605043. 
 
\bibitem{Gro98} 
L. Grover (1998):
A framework for fast quantum mechanical algorithms.
Proc. 30 Annual ACM Symp. on the Theory of Computing, 53--62, ACM Press New York. 
See also http://arXiv.org/abs/quant-ph/9711043. 

\bibitem{Gru99}
J. Gruska (1999):
Quantum Computing.
McGraw-Hill, London.


\bibitem{Hei93}
S. Heinrich (1993):
\newblock Random approximation in numerical analysis.
\newblock In: K.~D. Bierstedt, A.~Pietsch, W.~M. Ruess, and D.~Vogt, editors,
 Functional Analysis, 123 -- 171, Marcel {D}ekker.

\bibitem{Hei01a}
S. Heinrich (2001): 
Quantum summation with an application to integration. 
Journal of Complexity (to appear). See also 
http://arXiv.org/abs/quant-ph/0105116. 
 


\bibitem{Hei01b} 
S.\ Heinrich (2001): 
Quantum integration in 
Sobolev classes (in preparation).

\bibitem{HN01a} 
S.\ Heinrich and E.\ Novak (2001a): Optimal summation and integration by deterministic, 
randomized, and quantum algorithms. In:  K.-T. Fang, F.~J. Hickernell, and 
H.~Niederreiter, editors, Monte Carlo and Quasi-Monte Carlo Methods 2000, 
Springer-Verlag, Berlin (to appear),
see also http://arXiv.org/abs/quant-ph/0105114.

\bibitem{HN01b} 
S.\ Heinrich and E.\ Novak (2001b): On a problem in quantum summation.
Submitted to J. Complexity, see also http://arXiv.org/abs/quant-ph/0109038.

\bibitem{Man80}
Yu.\ I.\ Manin (1980): Computable and uncomputable (in Russian). Sovetskoye Radio,
Moscow.

\bibitem{Man99}
Yu.\ I.\ Manin (1999): Classical computing, quantum computing, 
and Shor's factoring algorithm. See http://arXiv.org/abs/quant-ph/9903008.

\bibitem{NW99} 
A. Nayak and F. Wu (1999):
The quantum query complexity of approximating the median and related statistics. 
STOC, May 1999, 384--393, 
see also http://arXiv.org/abs/quant-ph/9804066.

\bibitem{NC00}
M. A. Nielsen and I. L. Chuang (2000):
Quantum Computation and Quantum Information. Cambridge
University Press.

\bibitem{Nov88} 
E. Novak (1988):
Deterministic and Stochastic Error Bounds in Numerical Analysis. 
Lecture Notes in Mathematics {\bf 1349}, Springer. 

\bibitem{Nov01} 
E. Novak (2001):
Quantum complexity of integration.
J. Complexity {\bf 17}, 2--16.
See also http://arXiv.org/abs/quant-ph/0008124. 

\bibitem{Pit99} 
 A. O. Pittenger (1999):
Introduction to Quantum Computing Algorithms.
Birk\-h\"auser, Boston.

\bibitem{Rab80}
M.\ O.\ Rabin (1980): 
Probabilistic algorithm for testing primality. 
J. Number Theory 12, 128-138.

\bibitem{Sho94} 
P. W. Shor (1994):  Algorithms for quantum computation: 
Discrete logarithms and  factoring. Proceedings of the
35th Annual Symposium on Foundations of Computer Science,
IEEE Computer Society Press, Los Alamitos, CA, pp. 124--134. 
See also http://arXiv.org/abs/quant-ph/9508027. 

\bibitem{Sho98} 
P. W. Shor (1998):
Quantum computing. 
Documenta Mathematica, Extra Volume ICM 1998, I, 467--486.

\bibitem{Sho00}
P.\ W.\ Shor  (2000):
Introduction to quantum algorithms. 
See http://arXiv.org/abs/quant-ph/0005003.

\bibitem{Sol77}
 R.\ Solovay and V.\ Strassen (1977): 
A fast Monte-Carlo test for primality.
SIAM J. Comput. 6, 84-85.

\bibitem{TW01} 
J. F. Traub and H. Wo\'zniakowski (2001):
Path integration on a quantum computer. See
http://arXiv.org/abs/quant-ph/0109113.

\bibitem{TWW88} 
 J. F. Traub, G. W. Wasilkowski, and H. Wo\'zniakowski  (1988):
Information-Based Complexity. Academic Press. 

\end{thebibliography}
\end{document}